\documentclass[superscriptaddress,aps,prl,twocolumn]{revtex4-1}
\usepackage{graphicx}
\usepackage{bm}
\usepackage{color}
\usepackage{amsmath}

\newcommand{\sig}{ \mbox{\boldmath{$\sigma$}}}
\newcommand{\bet}{ \mbox{\boldmath{$\beta$}}}

\begin{document}

\title{Magnetic phase transitions in two-dimensional two-valley semiconductors with in-plane magnetic field}

\date{\today}

\author{Dmitry Miserev, Jelena Klinovaja, and Daniel Loss} 
\affiliation{Department of Physics, University of Basel, Klingelbergstrasse 82, CH-4056 Basel, Switzerland}

\begin{abstract} 
A two-dimensional electron gas (2DEG) in two-valley semiconductors has two discrete degrees of freedom given by the spin and valley quantum numbers.  We analyze the zero-temperature magnetic instabilities of two-valley semiconductors with SOI, in-plane magnetic field, and electron-electron interaction.
The interplay of an applied in-plane magnetic field and the SOI results in non-collinear spin quantization in different valleys. Together with the exchange intervalley interaction  this results in a rich phase diagram containing four non-trivial magnetic phases.
The negative non-analytic cubic correction to the free energy, which is  always present  in an interacting 2DEG, is responsible for  first order phase transitions.
Here, we show that non-zero  ground state values of the order parameters  can cut this cubic non-analyticity
and drive certain magnetic phase transitions  second order. 
We also find two tri-critical points at
zero temperature which together with the line of second order phase transitions constitute the quantum critical sector of the phase diagram. 
The phase transitions can be tuned externally by electrostatic gates  or by the in-plane magnetic field.
\end{abstract}

\maketitle

\begin{figure}[t]
	\includegraphics[width=0.9\columnwidth]{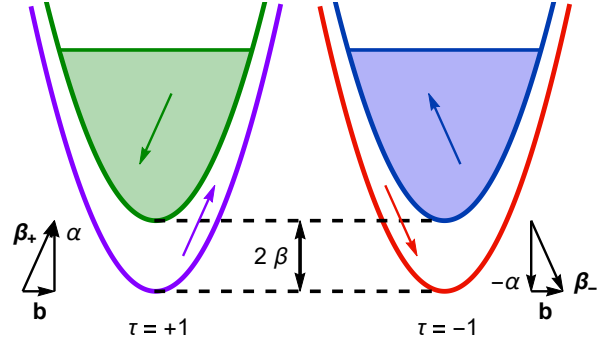}
	\vspace{-5pt}
	\caption{Electron spectrum
		in presence of the valley SOI 
		and the in-plane magnetic field,
		see Eq.~(\ref{spectrum}).
		Valleys are indicated by the index $\tau = \pm 1$.
		The spin degeneracy in each valley is lifted
		by the gap $2 \beta$, see Eq.~(\ref{spectrum}).
		The spin quantization axis 
		in $\tau = +1$ ($\tau = -1$)
		valley is directed along $\bet_+$ ($\bet_-$),
		see Eq.~(\ref{beta}),
		corresponding spin projections are 
		shown by arrows.
		Here we show 
		an example of equal filling of 
		green and blue bands,
		while purple and red bands are unfilled due to 
		the effect of electron-electron interaction.
		Such filling corresponds to the phase I',
		see Fig.~\ref{fig:phasediag},
		with $M_v = M_{\beta v} = 0$, $M_\beta \ne 0$,
		see Eq.~(\ref{mag}).
	}
	\label{fig:spec}
\end{figure}

{\it Introduction.}
Modern nanotechnology is mostly based on layered 
quantum materials where electrons or holes are confined within 
one layer which makes them effectively two-dimensional (2D)~\cite{dress}.
The 2D layers are typically represented by semiconductors 
such as GaAs, InAs, and InSb which are single valley materials
meaning that the electron energy has a single minimum in the Brillouin zone \cite{awschalom}.
Bulk Si and Ge have a valley degree of freedom, 
i.e. their bulk spectrum has several minima in the Brillouin zone (six for Si and four for Ge)~\cite{ando}.
In thin Ge films the valley degeneracy is lifted~\cite{ando},
while it survives in a Si 2D electron gas (2DEG).
This results in  qualitatively new physics in Si 2DEGs that has no analogues in single-valley 2DEGs.
For example, the valley degeneracy in Si 2DEGs allows for 
the singlet-triplet level crossing in 2D two-electron Si quantum dots \cite{miserdot},
an effect which is forbidden in  single-valley materials \cite{ashcroft}.
However, the spectrum of Si 2DEG still has a single minimum in 2D Brillouin zone.
A new class of 2D semiconductors with the electron spectrum containing two distinct minima (see Fig.~\ref{fig:spec})
is represented by monolayers of transition metal dichalcogenides (TMD)~\cite{wang,xu}.
These spectral minima are separated  by a wave vector $\bm q_0$, which is in order of the Brillouin zone momentum,
 $|\bm q_0| \sim 1/a_0$, $a_0$ being the lattice constant.

Magnetic instabilities in monolayers of TMDs 
were analyzed theoretically in Refs.~\cite{mukh,braz,donck,miserfer}.
Mixed ferromagnetic and valley polarized phases are 
predicted 
in the spin locking regime when the spin-orbit interaction (SOI)
is much larger than the Fermi energy $E_F$,
the regime that is realistic for 
the hole doped TMDs \cite{mukh,braz}
where the SOI gap is in order of few hundred meV \cite{rama,chei,zhao,ross}.
The opposite limit of small SOI compared to 
$E_F$ was considered in Refs.~\cite{donck,miserfer}
and is relevant for electron doped monolayers of TMDs,
especially for MoS$_2$
which has the smallest SOI gap $\sim 3 \,$meV 
among all monolayer TMDs~\cite{kadantsev,xiao,kosmider,liu,kli,korma,burkard}.
The exchange intervalley interaction and
the dynamical 
screening of the Coulomb interaction
are omitted in 
Ref.~\cite{donck}.
It was then shown subsequently~\cite{miserfer}
that these two ingredients 
have a dramatic effect on the magnetic phase diagram of TMDs.
In particular, finite exchange intervalley interaction
favors a ferromagnetic instability 
and biases out other possible magnetic orders.
The phase transition between the ferromagnetic and the paramagnetic ground states of
the 2DEG
is predicted to be of first order
due to the dynamical screening 
of the Coulomb interaction by gapless electron-hole fluctuations \cite{bkv}.
The theoretical results~\cite{miserfer}
agree well with recent experiments on 
electron doped monolayer MoS$_2$~\cite{roch,rochfirst,pisoni}.

Here, we study the 
effect of an applied in-plane magnetic field 
on the magnetic phase diagram of 
2D two-valley semiconductors such as 
electron doped monolayer TMDs, see Fig. \ref{fig:spec}.
This effect has 
not been studied theoretically so far
and leads to a rather rich magnetic phase diagram
allowing for the phase transitions to
be driven 
not only by the electron density,
which is tunable by electrostatic gates,
but also by the external magnetic field.
In this work we do not consider 
out-of-plane magnetic fields 
to avoid complications related to  Landau quantization.
We also assume that the SOI is much smaller 
than $E_F$ of the 2DEG in the normal phase.
Together with the intrinsic SOI
the in-plane magnetic field
leads to a non-collinear spin quantization
in different valleys
which are coupled by exchange intervalley interaction.
This breaks spin conservation 
which has a dramatic effect on the magnetic phase diagram.
Indeed, we show that four non-trivial 
magnetic orders are possible.
The phase transitions between the different phases can be driven by 
changing the electron density
and the external in-plane magnetic field.

In order to study the order of a phase transition,
we calculate the non-analytic cubic correction
to the free energy of the 2DEG
which comes from the dynamical screening of the 
Coulomb interaction and from the interaction vertex correction
due to gapless electron-hole fluctuations \cite{bkv}.
In case of a single 
magnetic order parameter (spin magnetization) 
the non-analytic cubic term is always negative \cite{belitz}
which results in a first order ferromagnetic phase transition 
at small temperatures~\cite{bkv,maslov}.
The finite temperature \cite{kirk}
or finite SOI \cite{zak1,zak2}
gap out the electron-hole continuum 
which is known to drive a second order magnetic phase transition~\cite{kb}.
However, the valley degree of freedom  in two-valley semiconductors results in 
three independent magnetic order parameters
that are coupled together 
via the non-analytic cubic correction.
The cubic correction is negative as in the case of
a single magnetic order parameter.
However, we show that 
certain phase transitions are of  second order
due to the interplay between all three order parameters 
coupled by the non-analytic cubic correction.
We also identify two tri-critical points on 
the zero-temperature phase diagram
that together with the line of 
second order phase transitions represent the quantum critical sector.

{\it Single-particle spectrum. }
The single-particle spectrum of a 2D two-valley 
semiconductor
can be described by the following effective Hamiltonian~\cite{korma}:
\begin{eqnarray}
&& H = \frac{\bm k^2}{2 m} - \alpha \sigma_z \tau - g \mu_B \bm B \cdot \frac{\sig}{2} , \label{ham}
\end{eqnarray}
where $\bm k = (k_x, k_y)$ is the in-plane momentum, 
$m$ the effective electron mass, 
$\alpha$ the valley SOI, 
$g$ the electron $g$-factor,
$\mu_B$ the Bohr magneton,
$\bm B = (B_x, B_y, 0)$ the in-plane magnetic field,
$\sig = (\sigma_x, \sigma_y, \sigma_z)$ the Pauli matrices
corresponding to the electron spin, and
$\tau = \pm 1$ labels the two valleys.
The valley SOI 
acts as an effective out-of-plane magnetic field 
taking opposite signs in different valleys.
Thus, the effective single-particle Hamiltonian 
(\ref{ham}) 
can be represented as a Zeeman Hamiltonian 
with a valley-dependent effective magnetic field $\bet_\tau$:
\begin{eqnarray}
&& H = \frac{\bm k^2}{2 m} - \bet_\tau \cdot \sig , \label{ham2} \\
&& \bet_\tau = (b_x, b_y, \alpha \tau), \, \bm b = g \mu_B \frac{\bm B}{2} . \label{beta}
\end{eqnarray}
The free-electron spectrum
described by Eq.~(\ref{ham2}), 
thus consists of two pairs of parabolic energy bands
(one pair per valley)
each of which is split by the corresponding effective magnetic field 
$\bet_\tau$,
see Fig.~\ref{fig:spec}:
\begin{eqnarray}
&& \varepsilon^\lambda_\tau (k) = \frac{k^2}{2 m} - \lambda \beta , \, \beta = |\bet_\tau| = \sqrt{\alpha^2 + b^2}, \label{spectrum}
\end{eqnarray}
where $\lambda = \pm 1$ is the eigenvalue
of the operator $\bet_\tau \cdot \sig / \beta$,
$k^2 = k_x^2 + k_y^2$, $b = |\bm b| = \sqrt{b_x^2 + b_y^2}$. 
As $|\bet_+| = |\bet_-| = \beta$, 
the single-particle spectrum is doubly degenerate.
The electron spin in valley $\tau$ is quantized along the direction $\bet_\tau / \beta$ of the effective magnetic field $\bet_\tau$,
see Fig.~\ref{fig:spec}.

{\it Magnetic order parameters. }
In absence of interactions 
 the  chemical potential 
is the same for all
single-particle bands, see Eq.~(\ref{spectrum}).
We refer to this ground state as  normal state.
Sufficiently strong electron-electron interactions
may result in unequal 
chemical potentials for different
single-particle bands.
For example,
strong electron-electron interaction in a single-valley 2DEG
results in unequal filling of the bands with different spin projections.
This leads to finite spin magnetization
even in absence of external magnetic fields.
This effect is known as Stoner instability \cite{stoner}.
In this work we investigate  magnetic instabilities 
in  2D two-valley semiconductors 
in the presence of an in-plane magnetic field, 
valley SOI, and electron-electron interactions.

The valley degree of freedom
together with the electron spin 
result in four single-particle bands,
see Fig.~\ref{fig:spec}.
Thus, we have four different electron densities $n^\lambda_\tau$, one per band,
which are related to the corresponding chemical potentials.
As the total electron density $n = \sum_{\lambda, \tau} n^\lambda_\tau $ is fixed by  electrostatic gates,
we have three independent magnetic order parameters
which we can change.
We make the following choice of the order parameters:
\begin{eqnarray}
M_\beta = \sum\limits_{\lambda, \tau} \frac{\lambda n^\lambda_\tau}{4 \nu},\ M_v = \sum\limits_{\lambda, \tau} \frac{\tau n^\lambda_\tau}{4 \nu},\ M_{\beta v} = \sum\limits_{\lambda, \tau} \frac{\lambda \tau n^\lambda_\tau}{4 \nu} ,
\label{mag}
\end{eqnarray}
where $\nu = m /(2 \pi)$ is the 2D density of states.
All order parameters have  dimension of energy.
The order parameter $M_v$ 
measures the valley magnetization, i.e.
the imbalance between the electron density in
different valleys.
The combination $M_\beta + M_{\beta v}$ ($M_\beta - M_{\beta v}$)
measures the spin magnetization in the $\tau = +1$ ($\tau = -1$) valley 
along the corresponding quantization axis given by the direction 
of the effective magnetic field 
$\bet_+$ ($\bet_-$), see Eq.~(\ref{beta}).
For example,
the filling shown in Fig.~\ref{fig:spec}
corresponds to $M_\beta \ne 0$, $M_v = M_{\beta v} = 0$.

{\it Interaction matrix elements. }
We separate the interaction matrix elements into two groups.
The first group corresponds to the direct Coulomb interaction, $v$, describing the electrostatic coupling of 
electric charges:
\begin{eqnarray}
&& v (q) = \nu \frac{2 \pi e^2}{\epsilon q} = \frac{1}{q a_B} ,
\end{eqnarray}
where we multiplied the Coulomb interaction by the 2D density of states $\nu = m /(2 \pi)$,
with $m$ being the effective mass,
$a_B = \epsilon / (m e^2)$ the effective Bohr radius,
and $\epsilon$  the dielectric constant.
At small momentum transfers $q \lesssim k_F$,
$k_F$ being the Fermi momentum, 
the 2D Coulomb  pole $1/q$ is screened by the electron-hole fluctuations
giving rise to the Thomas-Fermi screening momentum $q_{TF} \sim 1/a_B$,  see e.g.  Ref.~\cite{guinea}:
\begin{eqnarray}
&& v(q \lesssim k_F) \approx \frac{1}{(q + q_{TF}) a_B} . 
\label{TF}
\end{eqnarray}
The regime of strong electron-electron interaction corresponds to low densities such that $k_F a_B \ll 1$.
Thus, the $q a_B \lesssim k_F a_B \ll 1$ term is negligible compared to the $q_{TF} a_B \sim 1$ term in Eq.~(\ref{TF}),
which allows us to neglect the $q$-dependence in the direct interaction matrix element.

The second group of interaction matrix elements
corresponds to 
the exchange intervalley interaction $u$.
As the separation between the valleys is of the order of the Brillouin zone size,
such an interaction is extremely short ranged
and is mostly given by the on-site Hubbard term 
rather than by the long-range Coulomb interaction~\cite{guinea}.
Here we treat the direct, $v$, and the intervalley exchange, $u$, matrix elements 
as free phenomenological parameters.
We also assume that $u \ll v$, 
so we only account for the first order effects in $u$.

{\it Magnetic phase diagram. }
Magnetic instabilities are 
typically studied 
within the self-consistent Born (SCB) approximation
for the electron self-energy
that accounts for the Fermi liquid renormalizations
and the random phase approximation (RPA) 
for the dynamical screening of the Coulomb interaction \cite{maslov}.
Here, the RPA is motivated 
by a relatively large number of electron 
flavors due to the spin and valley quantum numbers \cite{maslov}.
In the static limit the RPA results in the Thomas-Fermi screening, see Eq.~(\ref{TF}).
Small dynamic corrections to the interaction
which are not shown in Eq.~(\ref{TF})
are responsible for the cubic non-analyticity in the free energy~\cite{maslov}.

Neglecting the dynamical screening (i.e. treating $u$ and $v$ as constants)
and applying the SCB approximation, 
we calculate the free energy of the two-valley 2DEG, 
see the Supplementary Material (SM)~\cite{SM}:
\begin{eqnarray}
&& F^{SCB} = 2 \nu \left[ a_\beta M_\beta^2 + a_v M_v^2 + a_{\beta v} M_{\beta v}^2 - 2 \beta M_\beta \right] ,
\label{SCB}
\end{eqnarray}
where we subtracted the energy of 
the normal state.
Here we introduced the Landau parameters:
\begin{eqnarray}
&& a_\beta = 1 - v - \zeta u, \label{ab}\\
&& a_v = 1 - v + u, \label{av} \\
&& a_{\beta v} = 1 - v + \zeta u, \label{abv}
\end{eqnarray}
where 
\begin{eqnarray}
&& \zeta = \frac{b^2 - \alpha^2}{\beta^2} = \frac{b^2 - \alpha^2}{b^2 + \alpha^2} .
\end{eqnarray}
All order parameters are assumed to be much smaller than the Fermi energy $E_F = n /(4 \nu)$ in the normal phase, $n$ is the total density.
The order parameter 
$M_\beta$ is non-zero even in
the normal phase 
due to the $\beta M_\beta$ term, see Eq.~(\ref{SCB}), 
that results in small linear 
in $\beta \ll E_F$ polarization $M_\beta \propto \beta$.
In what follows we assume that the spontaneous magnetization
is always much larger than the finite field polarization $\propto \beta$, and, thus, we neglect the $\beta M_\beta$ term.
We only keep in mind that finite $\beta$ 
explicitly breaks the 
$M_\beta \to -M_\beta$ symmetry of the free energy,
and, thus,
is responsible for the sign of $M_\beta$,
which, however, is not important for our consideration.

\begin{figure}[t]
	\includegraphics[width=\columnwidth]{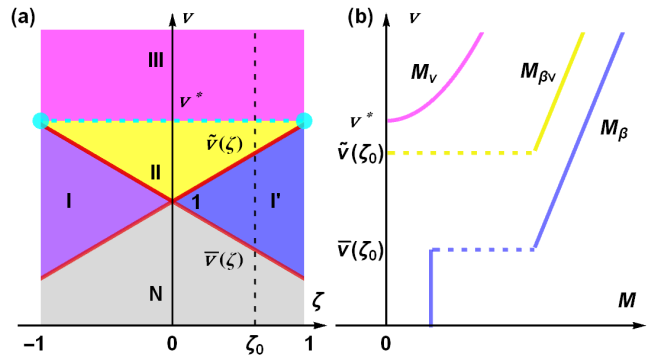}
	\caption{(a) The magnetic phase diagram.
		The following phases are shown:
		the normal phase $N$ (gray) with
		no condensed order parameters;
		the homogeneous phases I (purple) and I' (blue) 
		with single condensed order parameter $M_{\beta v}$
		and $M_{\beta}$, respectively;
		the phase II (yellow) with two condensed order parameters 
		$M_\beta$ and $M_{\beta v}$;
		the phase III (magenta) with non-zero ground state value of all three magnetic order parameters.
		The first (second) order phase transitions are shown by 
		solid red (dashed cyan) lines.
		Two tri-critical points are indicated by cyan dots.
		The lines of phase transitions are given by Eqs.~(\ref{vbar})--(\ref{vstar}).
		(b) Schematic dependence of the order parameters along the cut $\zeta = \zeta_0$ on the phase diagram. Finite jump of $M_\beta$ and $M_{\beta v}$ reflects the first order phase transitions, continuous dependence of $M_v$ on $v$ is due to the second order phase transition. Non-zero $M_\beta \propto \beta$ at $v < \bar v(\zeta_0)$ is due to finite external field $\beta$, see $\beta M_\beta$ term in Eq.~(\ref{SCB}).}
	\label{fig:phasediag}
\end{figure}

Let us consider the case $1 > \zeta > 0$.
In this case $a_v > a_{\beta v} > a_{\beta}$ at any $u > 0$ and $v > 0$.
Starting from the normal phase at $v = 0$ and increasing the value of $v$
we first find the phase transition for the 
order parameter $M_\beta$ when $a_\beta = 0$, 
which corresponds to the following critical value of $v = \bar v(\zeta)$:
\begin{eqnarray}
&& \bar v (\zeta) = 1 - \zeta u . 
\label{vbar}
\end{eqnarray} 
This corresponds to the phase transition from the normal 
phase (N) to the phase I' 
with condensed order parameter $M_\beta \ne 0$,
see Fig.~\ref{fig:phasediag}(a).
Upon further increasing the interaction parameter $v$,
we find an instability for the $M_{\beta v}$ order parameter 
at $v = \tilde v(\zeta)$ at which $a_{\beta v} = 0$:
\begin{eqnarray}
&& \tilde v(\zeta) = 1 + \zeta u .
\label{vzeta}
\end{eqnarray}
This corresponds to the phase transition from the ordered phase I'
with the only condensed order parameter $M_\beta \ne 0$
to the phase II with $M_\beta \ne 0$, $M_{\beta v} \ne 0$, $M_v = 0$.
Increasing $v$ even further, we reach the 
instability for the valley order parameter $M_v$ 
when $a_v = 0$.
This happens at $v = v^*$:
\begin{eqnarray}
&& v^* = 1 + u . \label{vstar}
\end{eqnarray} 
This corresponds to the phase transition from the phase II 
to the phase III where all magnetic order parameters are condensed $M_\beta \ne 0$, $M_{\beta v} \ne 0$, $M_v \ne 0$.
Further increase of the matrix element $v$ does not result 
in additional instabilities within the SCB approximation,
the system stays in the phase III at any $v > v^*$.

	Similar analysis can be done for $\zeta < 0$.
	Upon increasing $v$ from $v = 0$ we first find 
	the instability for $M_{\beta v}$ 
	at $v = \tilde v(\zeta) = 1 + \zeta u$,
	then the instability for $M_\beta$ 
	at $v = \bar v(\zeta) = 1 - \zeta u$,
	and then the instability for $M_v$
	at $v = v^* = 1 + u$.
	This results in the phase diagram shown in Fig.~\ref{fig:phasediag}(a).

{\it Order of phase transitions. }
Normally, the magnetic phase transitions
are first order due to the
non-analytic cubic 
correction to the free energy \cite{pt}
that originates from the dynamical screening of the 
Coulomb interaction and from the interaction vertex correction \cite{pepin}
that are neglected within the SCB approximation.
In the SM~\cite{SM} we calculate the cubic correction within  
second order perturbation theory neglecting the 
$u^2$ term because we assume $u \ll v \sim 1$:
\begin{eqnarray}
&& F^{cub} =  - \frac{2 \nu v^2}{3 E_F} \left[\left(1 - \frac{2 u}{v} \frac{b^2}{\beta^2}\right) f\left(M_v, M_{\beta v}\right) \right.  \label{cub}\\
&& \left.  + \left(1 - \frac{2 u}{v} \frac{\alpha^2}{\beta^2}\right) f\left(M_v, M_{\beta}\right) + f \left(M_\beta, M_{\beta v}\right) \right], \nonumber
\end{eqnarray}
where 
$f(x, y) = f(y, x)$ is a symmetric non-analytic cubic function of two variables:
\begin{eqnarray}
&& f(x, y) = |x|^3 + 3 |x| y^2, \, |x| \ge |y| .
\label{f}
\end{eqnarray}
Importantly, $f(x, y)$ is non-analytic only with respect to the largest in absolute value argument.
This means that if all three order parameters are condensed,
see phase III in Fig.~\ref{fig:phasediag}(a), then
the non-analyticity for the smallest order parameter is cut by larger ground state values of two other order parameters, 
see Eqs.~(\ref{cub}) and (\ref{f}).
As the Landau parameter $a_v$ is always the largest,
see Eqs.~(\ref{ab})--(\ref{abv}),
then $M_v$ order parameter is more suppressed compared to $M_\beta$ and $M_{\beta v}$
i.e. $|M_v| < |M_\beta|$ and $|M_v| < |M_{\beta v}|$ in phase III.
Therefore, there is no cubic in $M_v$ term in the free energy.
The absence of $|M_v|^3$ term 
results in the second order phase transition between phases II and III.
This is illustrated in Fig.~\ref{fig:phasediag}(b) by the continuous dependence 
of $M_v$ on $v$.

In all magnetic phases
apart from the phase III $M_v = 0$, see Fig.~\ref{fig:phasediag}(a).
Therefore, in these phases
$f(M_v, M_{\beta v}) = |M_{\beta v}|^3$, 
 $f(M_v, M_{\beta}) = |M_{\beta}|^3$,
i.e. the cubic non-analyticity for $M_\beta$ and $M_{\beta v}$ 
is not cut in Eq.~(\ref{cub}).
As negative cubic non-analyticity necessarily leads to a first
order phase transition \cite{pt},
all other magnetic phase transitions are of the first order
which is illustrated in Fig.~\ref{fig:phasediag}(b)
by the finite jumps of corresponding order parameters at the phase transition.
For example, the first order ferromagnetic phase transition 
between normal phase and phase I
at zero applied magnetic field, $\zeta = -1$,
has been predicted in Ref.~\cite{miserfer}.

Due to finite external field $\beta$, $M_\beta$ takes non-zero value $M_\beta \propto \beta$
even in normal phase
which can be an issue for the experimental identification of phase I'.
However, the discontinuity of $M_\beta$ 
at the first order phase transition between 
normal phase and phase I' 
unambiguously signals the spontaneous symmetry breaking, see Fig.~\ref{fig:phasediag}(b).

The zero temperature phase diagram 
shown in Fig.~\ref{fig:phasediag}(a)
contains two tri-critical points
located at $v = v^*$ and $\zeta = \pm 1$
where the first and second order phase transition lines meet.
The tri-critical points correspond to 
either zero magnetic field or zero SOI
when all magnetic order parameters are Ising.
We point out here that 
these tri-critical points have been overlooked in the previous theoretical studies \cite{donck,miserfer}.
Second order phase transitions,
see dashed cyan line $v = v^*$ 
in Fig.~\ref{fig:phasediag}(a),
and the tri-critical points 
result 
in quantum critical states that
are characterized by  
emergent long range order
and divergent susceptibilities \cite{sachdev}.
Even though we predict these quantum critical points,
we cannot describe them quantitatively
within the mean field treatment that we use in this paper 
due to the increasingly important effect of
the fermion fluctuations in the vicinity of such quantum critical points,
for more information see Ref.~\cite{mvojta}.

{\it Conclusions. }
In our study we predict a very rich
magnetic phase diagram 
in 2D two-valley semiconductors with intrinsic valley SOI, 
in-plane magnetic field, and electron-electron
interaction, see Fig.~\ref{fig:phasediag}(a).
Phase transitions between different phases 
can be driven by the electron density 
(varying the interaction parameter $v$)
or by the external in-plane magnetic field.
In spite of the cubic non-analytic corrections to the free energy that 
favors  first order phase transitions,
we showed that the phase transition between 
phases II and III, see Fig.~\ref{fig:phasediag}(a),
is second order.
Together with two tri-critical points,
the line of second order 
phase transitions constitute
the quantum
critical sector on the zero-temperature phase diagram.

{\it Acknowledgments.}
This work was supported by the Georg H. Endress foundation, the Swiss National Science Foundation, and NCCR QSIT. This project received funding from the European Union's Horizon 2020 research and innovation program (ERC Starting Grant, grant agreement No 757725).

\newpage
\onecolumngrid
\
\begin{center}
	\large{\bf Supplemental Material for ``Magnetic phase transitions in two-dimensional two-valley semiconductors with in-plane magnetic field  '' \\}
\end{center}
\begin{center}
	Dmitry Miserev, Jelena Klinovaja, and Daniel Loss\\
	{\it Department of Physics, University of Basel, Klingelbergstrasse 82, CH-4056 Basel, Switzerland}\\

\end{center}

\setcounter{equation}{0}
\setcounter{figure}{0}

\renewcommand{\theequation}{S\arabic{equation}}
\setcounter{equation}{0}  
\renewcommand{\thefigure}{S\arabic{figure}}
\setcounter{figure}{0}  
\twocolumngrid

In the Supplemental Material (SM) 
we provide detailed calculations of the 
grand canonical potential $\Omega$
and the free energy $F$
within the self-consistent Born (SCB) approximation
for the electron self-energy 
and the random phase approximation (RPA) for the dynamically screened 
Coulomb potential.
We first start from the non-interacting 2DEG.
Then we account for the Fermi liquid renormalizations
via the SCB approximation 
accounting for the Thomas-Fermi screening of the
Coulomb interaction.
Next we calculate the effect of the dynamical screening 
and the interaction vertex correction 
within second order perturbation theory
which results in the 
cubic non-analyticity in the grand canonical potential.
After that we perform the Legendre transformation
in order to obtain the free energy 
which we analyze in the main text
in terms of magnetic instabilities.

\section{Non-interacting 2DEG}

Here we calculate the grand canonical 
potential for a non-interacting two-valley 2DEG
with valley SOI and  in-plane 
magnetic field.
The effective single-particle Hamiltonian 
is given in Eqs.~(1) and (2) in the main text.
The electron spectrum is given by 
the twofold degenerate parabolic bands, see Eq.~(4)
in the main text.
Here, we also require the electron spinors 
which explicitly show non-collinear spin quantization in different valleys:
\begin{eqnarray}
&& |\lambda \tau \rangle = \frac{1}{\sqrt{2 \beta (\beta - \alpha \lambda \tau)}} {b_x - i b_y \choose \lambda (\beta - \alpha \lambda \tau)}, \label{spinor}
\end{eqnarray}
where $\lambda = \pm 1$ is the eigenvalue of 
$\bet_\tau \cdot \sig / \beta$, 
$\tau = \pm 1$ labels the valley,
$\alpha$ is the valley SOI,
$\bm b$, $\bet_\tau$, and $\beta$ are defined in Eq.~(3) in the main text.
In what follows we also need the spinor overlap matrix elements:
\begin{eqnarray}
&& M^{\lambda' \lambda}_{\tau \tau} = \langle \lambda' \tau| \lambda \tau \rangle = \delta^{\lambda' \lambda}, \label{m1}\\
&& M^{\lambda' \lambda}_{-\tau \tau} = \langle \lambda' -\tau| \lambda \tau \rangle = \frac{b}{\beta} \delta^{\lambda' \lambda} + \frac{\alpha \lambda \tau}{\beta} \delta^{-\lambda' \lambda} , \label{m2}
\end{eqnarray}
where $\delta^{ab}$ is the Kronecker symbol.
The electron Matsubara Green function 
of the non-interacting 2DEG
is diagonal in the basis of spinors $|\lambda \tau\rangle$, 
see Eq.~(\ref{spinor}):
\begin{eqnarray}
&& G (K) = \sum\limits_{\lambda, \tau} |\lambda \tau \rangle \langle \lambda \tau| g_\tau^\lambda (K) ,  \label{green}\\
&& g_\tau^\lambda (K) = \frac{1}{i \omega - \varepsilon_0 (k) + \lambda \beta + \mu^\lambda_\tau}\,\, ,
\label{g}
\end{eqnarray}
where $K = (\bm k, \omega)$ is the ``relativistic'' notation for the momentum $\bm k = (k_x, k_y)$ and the Matsubara frequency $\omega$, $\varepsilon_0 (k) = k^2 / 2 m$,
$\mu^\lambda_\tau$ are the chemical potentials. 
It is convenient to introduce the notation
for chemical potentials $\tilde \mu^\lambda_\tau$ that 
are calculated from the bottom of the corresponding band:
\begin{eqnarray}
\tilde \mu^\lambda_\tau = \mu^\lambda_\tau + \lambda \beta .
\label{mutilde}
\end{eqnarray}

The non-interacting part of the grand canonical potential (per unit area) is given by the following expression:
\begin{eqnarray}
\Omega_0 = \sum\limits_{\lambda, \tau} \sum_K \ln \left[g^\lambda_\tau (K)\right] ,
\label{O0}
\end{eqnarray} 
where $\sum\limits_K = T \sum\limits_\omega \int d \bm k / (2 \pi)^2$ ($\hbar=k_B=1$).
Evaluating the sum in Eq.~(\ref{O0}), we get
\begin{equation}
\Omega_0 = - \frac{\nu}{2} \sum \limits_{\lambda, \tau} (\tilde \mu^\lambda_\tau)^2 ,
\label{om0}
\end{equation}
where $\nu = m /2 \pi$ is the 2D density of states and
$m$  the effective mass.

\section{SCB approximation}

In this section we calculate the effect 
of the Fermi liquid renormalization 
that we account for via the SCB approximation.
Here we only account for the Thomas-Fermi screening
of the Coulomb interaction
that leads to effectively short-range 
interaction matrix elements $v$ 
and $u$,
$v$ being the direct Coulomb interaction,
$u \ll v$ the exchange intervalley interaction.
For more information see 
the discussion in the main text
after Eq.~(6).

\begin{figure}[t]
	\includegraphics[width=0.99\columnwidth]{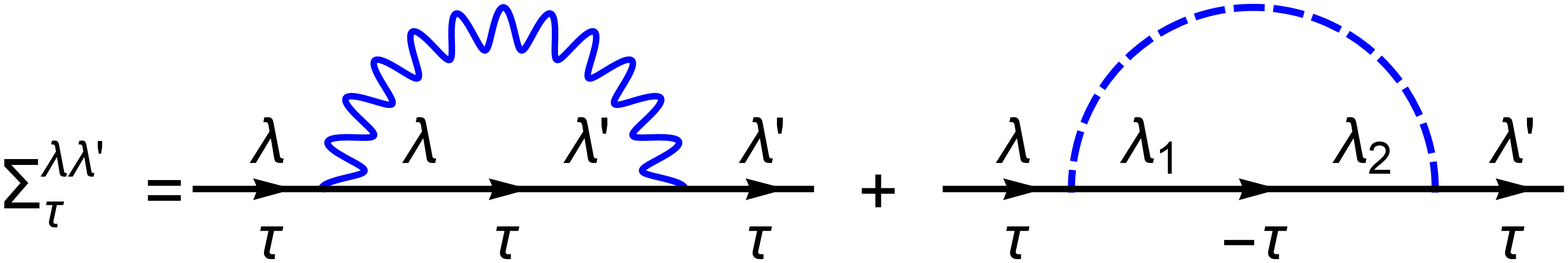}
	\caption{The SCB approximation 
		for the self-energy.
		Blue wavy (dashed) lines correspond to the $v$ ($u$) components of the effective interaction. 
		The black solid lines correspond to the electron Green function 
		self-consistently dressed by the self-energy,
		see Eq.~(\ref{gbar}).}
	\label{fig:scb}
\end{figure}

Within the SCB approximation,
the self-energy is given by the 
diagrams in Fig.~\ref{fig:scb}:
\begin{eqnarray}
&& \Sigma^{\lambda \lambda'}_\tau (K) = - \frac{v}{\nu} \sum\limits_P \bar g^{\lambda \lambda'}_\tau(P) \nonumber \\
&& - \frac{u}{\nu} \sum\limits_{\lambda_1, \lambda_2} \sum\limits_P M^{\lambda \lambda_1}_{\tau -\tau} M^{\lambda_2 \lambda'}_{-\tau \tau} \bar g^{\lambda_1 \lambda_2}_{-\tau} (P), 
\label{sigscb}
\end{eqnarray}
where the interaction is taken as contact interaction due to 
the Thomas-Fermi screening and the
matrix elements $M^{\lambda_1 \lambda_2}_{\tau_1 \tau_2}$
are given by Eqs.~(\ref{m1}) and (\ref{m2}).
The Green function $\bar g^{\lambda\lambda'}_\tau (P)$ is self-consistently dressed by the self-energy:
\begin{eqnarray}
&& \bar g_\tau (\omega, \bm k) = \left(i \omega  - \varepsilon_0(k) + \tilde \mu_\tau - \Sigma_\tau\right)^{-1} ,
\label{gbar}
\end{eqnarray}
where $\bar g_\tau (\omega, \bm k)$ is a $2\times 2$ 
matrix in the spin space,
$\tilde \mu_\tau^{\lambda \lambda'} = \tilde \mu_\tau^\lambda \delta^{\lambda \lambda'}$,
$\tilde \mu_\tau^\lambda$ is given by Eq.~(\ref{mutilde}).
This is clear from Eq.~(\ref{sigscb})
that the exchange intervalley scattering $u$ 
together with the non-collinearity 
of spins in different valleys
breaks the valley spin conservation.
The valley index of the single-particle Green function
is conserved due to the momentum conservation
because different valleys correspond 
to different momentum sectors.

Due to the contact approximation of the 
electron-electron interaction by constant matrix elements $v$ and $u$,
the self-energy, see Eq.~(\ref{sigscb}),
does not depend on frequency and momentum.
Therefore,
the sum over momenta $P$ in Eq.~(\ref{sigscb})
can be evaluated even for the dressed Green function:
\begin{eqnarray}
\frac{1}{\nu}\sum \limits_P \bar g^{\lambda \lambda'}_\tau (P) = \tilde \mu^\lambda_\tau \delta^{\lambda \lambda'} - \Sigma^{\lambda \lambda'}_\tau \equiv S^{\lambda \lambda'}_\tau ,
\label{s}
\end{eqnarray}
where we introduced a new notation for convenience.
Using Eq.~(\ref{s}), we simplify the
SCB Eq.~(\ref{sigscb}) 
to the following matrix equation:
\begin{eqnarray}
&& S_\tau = \tilde \mu_\tau + v S_\tau + u M_{\tau - \tau} S_{-\tau} M_{-\tau \tau} ,
\label{stau}
\end{eqnarray}
where $\tilde \mu^{\lambda \lambda'}_\tau = \tilde \mu^\lambda_\tau \delta^{\lambda \lambda'}$,
and the matrices $M_{-\tau \tau}$ and $M_{\tau -\tau}$
have spin matrix elements given by Eq.~(\ref{m2}).
Changing $\tau \to -\tau$ in Eq.~(\ref{stau}),
we get the second equation connecting $S_\tau$ and $S_{-\tau}$.
The solution of Eq.~(\ref{stau}) is the following:
\begin{eqnarray}
&& S_\tau = \frac{(1 - v) \tilde \mu_\tau + u M_{\tau -\tau} \tilde \mu_{-\tau} M_{-\tau \tau}}{(1 - v)^2 - u^2} .
\label{ssol}
\end{eqnarray}

In order to derive the grand canonical potential,
we first calculate the following auxiliary energy functional \cite{abrikosov}:
\begin{eqnarray}
&& \Phi = \sum\limits_{\lambda, \lambda', \tau} \sum\limits_P \Sigma^{\lambda \lambda'}_\tau \bar g^{\lambda' \lambda}_\tau (P) = \nu \sum\limits_{\lambda, \lambda', \tau} \Sigma^{\lambda \lambda'}_\tau S^{\lambda' \lambda}_\tau ,
\label{phi}
\end{eqnarray}
where we used Eq.~(\ref{s}) to calculate the sum over $P$.
Substituting Eq.~(\ref{ssol}) into Eq.~(\ref{phi}),
we find the $\Phi$ potential:
\begin{eqnarray}
&& \Phi(v, u) = - \nu \Phi_1(v, u) \sum\limits_{\lambda,\tau} \left(\tilde \mu^\lambda_\tau\right)^2 \nonumber \\
&& - \nu \Phi_2(v, u) \sum\limits_{\lambda, \tau} \left(\tilde \mu^\lambda_\tau \tilde \mu^{\lambda}_{-\tau} \frac{b^2}{\beta^2} + \tilde \mu^\lambda_\tau \tilde \mu^{-\lambda}_{-\tau} \frac{\alpha^2}{\beta^2}\right) ,
\end{eqnarray}
where we introduced the following prefactors:
\begin{eqnarray}
&& \Phi_1(v, u) = \frac{v (1 - v)^2 + u^2 (2 - v)}{\left[(1 - v)^2 - u^2\right]^2} , \label{phi1} \\
&& \Phi_2(v, u) = \frac{u(1 - v^2 + u^2)}{\left[(1 - v)^2 - u^2\right]^2} . \label{phi2}
\end{eqnarray}

The grand canonical potential is connected to the $\Phi$ 
potential through the following identity \cite{abrikosov}:
\begin{eqnarray}
&& \Omega = \Omega_0 + \frac{1}{2} \int\limits_0^1 \frac{ds}{s} \Phi(s v, s u) ,
\label{OmPhi}
\end{eqnarray}
where $\Omega_0$ is given by Eq.~(\ref{om0}).
Thus, we have to calculate the following elementary integrals:
\begin{eqnarray}
&& \int\limits_0^1 \frac{ds}{s} \Phi_1(s v, s u) = \frac{v (1 - v) + u^2}{(1 - v)^2 - u^2} , \\
&& \int\limits_0^1 \frac{ds}{s} \Phi_2(s v, s u) = \frac{u}{(1 - v)^2 - u^2} .
\end{eqnarray}
Substituting these integrals back into
Eq.~(\ref{OmPhi}), we find the grand canonical 
potential within the SCB approximation:
\begin{eqnarray}
&& \Omega^{SCB} = -\frac{\nu}{2} \frac{1}{(1-v)^2 - u^2} \sum\limits_{\lambda, \tau} \left[(1 - v) (\tilde \mu^\lambda_\tau)^2 \right. \nonumber \\
&& \left. + u \left( \frac{b^2}{\beta^2} \tilde \mu^\lambda_\tau \tilde \mu^\lambda_{-\tau} + \frac{\alpha^2}{\beta^2} \tilde \mu^\lambda_\tau \tilde \mu^{-\lambda}_{-\tau} \right) \right] . \label{omscb}
\end{eqnarray}

\section{Non-analytic cubic correction}

\begin{figure}[t]
	\includegraphics[width=\columnwidth]{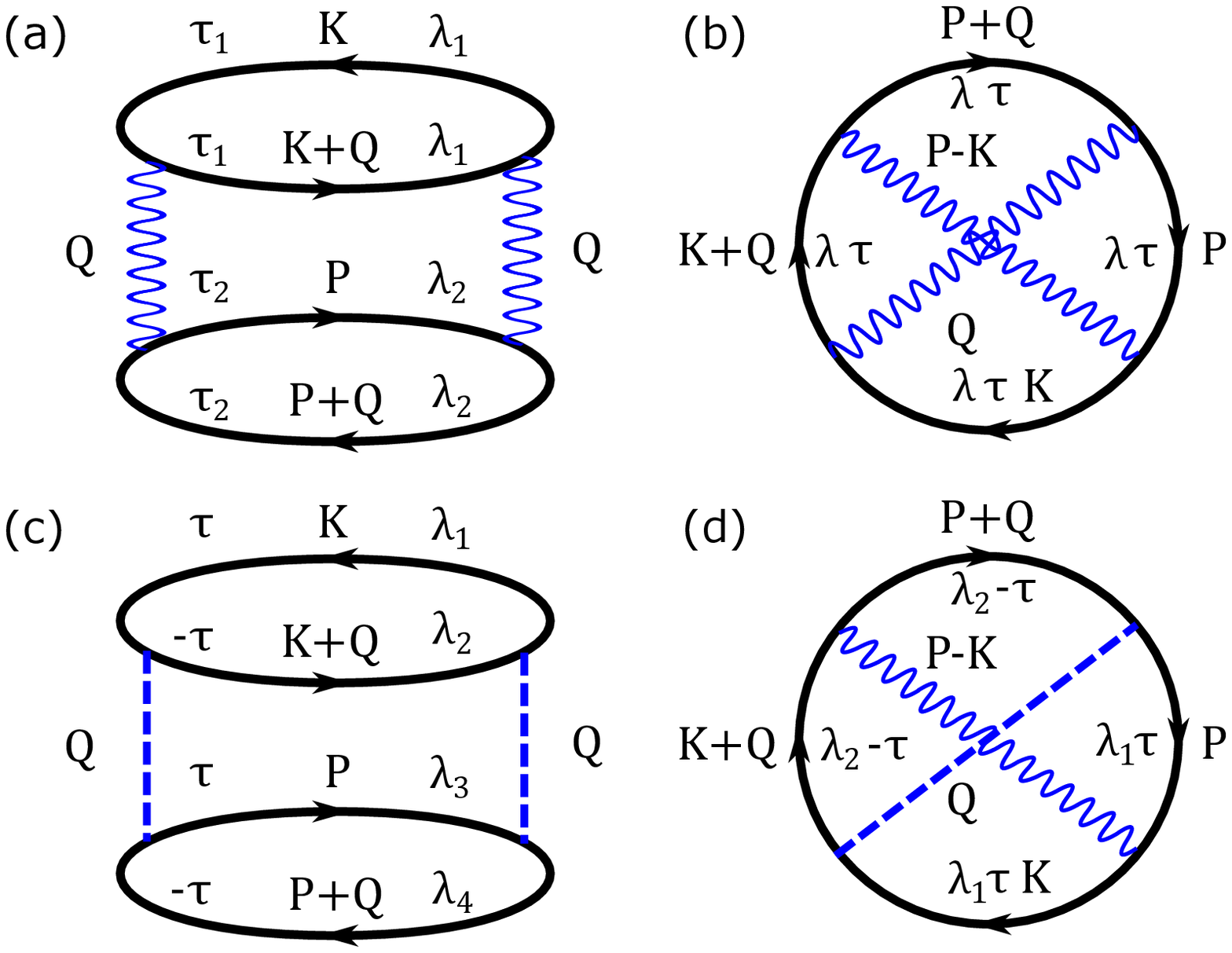}
	\caption{Second order diagrams for the grand canonical potential $\Omega$ beyond the SCB approximation. 
		Blue wavy (dashed) line corresponds to the $v$ ($u$) component of the effective interaction. 
		The black solid lines correspond to the bare electron Green function, see Eq.~(\ref{g}).}
	\label{fig:diag}
\end{figure}

Within the SCB approximation 
we neglected the dynamic dependence of the Coulomb interaction 
and the interaction vertex correction.
Here we account for these effects 
within second order perturbation expansion.
It is well known that these effects
give rise to the non-analytic cubic terms in the $\Omega$ potential \cite{belitz,bkv,kirk,kb}.
The diagrams that contribute to the cubic non-analyticity are shown in Fig. \ref{fig:diag}.
In order to make the formal expressions corresponding to these diagrams more compact,
we introduce the particle-hole bubble:
\begin{equation}
\Pi^{\lambda \lambda'}_{\tau \tau'}(Q) = \frac{1}{\nu} \sum_K g^\lambda_\tau(K) g^{\lambda'}_{\tau'}(K + Q) . \label{bubble}
\end{equation} 
The diagrams in Fig.~\ref{fig:diag}(b),(d) do not appear 
in the SCB series and represent the 
renormalizations of the interaction vertex.
However, the other two diagrams
are partially
accounted within the approximation that we have done
in the previous section 
because we included the Thomas-Fermi screening
of the Coulomb interaction.
In order to avoid the double counting,
we subtract the static part of the particle-hole bubbles
that is responsible for the Thomas-Fermi screening
from the diagrams in Fig.~\ref{fig:diag}(a),(c).
In other words, we only account for the dynamic part of the screened Coulomb interaction
in Fig.~\ref{fig:diag}(a),(c).
As we are interested in the lowest order perturbation correction
to the cubic non-analyticity,
we also neglect the Fermi liquid renormalizations,
i.e. the Green functions in the 
diagrams in Fig.~\ref{fig:diag} 
are assumed to be bare, see Eq.~(\ref{g}).

The diagrams in Fig. \ref{fig:diag} 
have the following analytic representations:
\begin{eqnarray}
\!\!\!\!\!\!\!\!\! \Omega_a & = & -\frac{v^2}{4} \sum\limits_{\tau_1,\tau_2} \sum\limits_{\lambda_1, \lambda_2} \sum\limits_Q \Pi^{\lambda_1 \lambda_1}_{\tau_1 \tau_1} (Q) \Pi^{\lambda_2 \lambda_2}_{\tau_2 \tau_2} (Q), \label{xia} \\
\!\!\!\!\!\!\!\!\! \Omega_b  & = & \frac{v^2}{4} \sum\limits_{\tau, \lambda} \sum\limits_Q \left[\Pi^{\lambda \lambda}_{\tau \tau} (Q)\right]^2, \label{xib} \\
\!\!\!\!\!\!\!\!\! \Omega_c & = & -\frac{u^2}{4} \! \!\sum\limits_{\tau, \lambda_i, Q} \!\!|M^{\lambda_1 \lambda_2}_{\tau - \tau} M^{\lambda_3 \lambda_4}_{\tau - \tau}|^2 \Pi^{\lambda_1 \lambda_2}_{\tau -\tau} (Q) \Pi^{\lambda_3 \lambda_4}_{\tau -\tau} (Q) , \label{xic} \\
\!\!\!\!\!\!\!\!\! \Omega_d & = & 2 \frac{u v}{4} \sum\limits_{\tau, \lambda_1, \lambda_2} \sum\limits_Q |M^{\lambda_1 \lambda_2}_{\tau -\tau}|^2 \left[\Pi^{\lambda_1 \lambda_2}_{\tau -\tau} (Q) \right]^2, \label{xid}
\end{eqnarray}
where $\nu = m/ 2 \pi$ is the 2D density of states,
$m$ is the effective mass. 
We note  the additional factor of $2$ in $\Omega_d$.
This is because $\Omega_d$ is represented by two diagrams: one is shown in Fig.~\ref{fig:diag}(d) and the other is different by swapping dashed and wavy interaction lines.

The non-analyticity in $\Omega$ comes from the non-analyticities of the particle-hole bubble $\Pi^{\lambda \lambda'}_{\tau \tau'}(Q)$, $Q = (\bm q, \varepsilon)$, which are known as  Landau damping at $Q \approx 0$ and  Kohn anomaly at 
$q = 2 k_F$, $\varepsilon = 0$, and $k_F$ is the Fermi momentum \cite{maslov}. 
First, we find the Landau damping contribution using the following approximation of the particle-hole bubble at small $Q = (\bm q, \varepsilon)$,
see e.g.  Refs. \cite{zak1,zak2}:
\begin{eqnarray}
&&\!\!\!\!\!\!\!\!\! \Pi_{\tau \tau'}^{\lambda \lambda'}(\bm q, \varepsilon) \approx \frac{|\varepsilon|}{\sqrt{\left(\varepsilon - i \Delta^{\lambda \lambda'}_{\tau \tau'}\right)^2 + \left(v_F q\right)^2}} - 1, \label{piapprox}
\end{eqnarray}
where $v_F = k_F / m$ is the Fermi velocity and 
\begin{equation}
\Delta^{\lambda \lambda'}_{\tau \tau'} = \tilde \mu^\lambda_\tau - \tilde \mu^{\lambda'}_{\tau'}.
\end{equation} 
It is assumed here that $\Delta^{\lambda \lambda'}_{\tau \tau'} \ll \tilde \mu^\lambda_\tau, \tilde \mu^{\lambda'}_{\tau'} $.
Then, the calculation of any sum which is quadratic with respect to the particle-hole bubble is straightforward and can be found e.g. in Ref. \cite{miserfer}:
\begin{eqnarray}
\!\!\!\!\!\!\!\!\!\!\!\!\sum_Q \!\! \vphantom{|}^{L} \Pi^{\lambda_1 \lambda_2}_{\tau_1 \tau_2}(Q) \Pi^{\lambda_3 \lambda_4}_{\tau_3 \tau_4}(Q) \!=\! \frac{\nu T^3}{4 E_F}\! \mathcal{F}\!\left(\frac{\Delta^{\lambda_1 \lambda_2}_{\tau_1 \tau_2} + \Delta^{\lambda_3 \lambda_4}_{\tau_3 \tau_4}}{2 T}\right)\!,
\label{Ssum2}
\end{eqnarray}
where the index $L$ of the sum indicates that $Q$ is in the vicinity of the Landau damping point.
The function $\mathcal{F} (z)$ has the following integral representation:
\begin{equation}
\mathcal{F}(z) = \int\limits_0^{z} dx \, x^2 \coth\left(\frac{x}{2}\right).
\label{F}
\end{equation}
At small temperature $T \to 0$ the argument of $\mathcal{F}$ in Eq.~(\ref{Ssum2}) is large and one can use the asymptotic expansion:
\begin{eqnarray}
\mathcal{F}(z) = \frac{|z|^3}{3} + 4 \zeta(3) + O(e^{-|z|}), \quad |z| \gg 1 ,
\label{expansion}
\end{eqnarray}
where $\zeta(z)$ is the Riemann $\zeta$-function.
The function $\mathcal{F}(z)$ results in the cubic non-analyticity of the $\Omega$ potential.

The Kohn anomaly contribution can be reduced to the Landau damping with the help of a trick used in Refs. \cite{maslov,miserfer}:
\begin{equation}
\sum_Q \!\! \vphantom{|}^{K} \Pi^{\lambda_1 \lambda_2}_{\tau_1 \tau_2}(Q) \Pi^{\lambda_3 \lambda_4}_{\tau_3 \tau_4}(Q) \! = \frac{\nu T^3}{4 E_F} \mathcal{F}\left(\frac{\Delta^{\lambda_1 \lambda_3}_{\tau_1 \tau_3} + \Delta^{\lambda_2 \lambda_4}_{\tau_2 \tau_4}}{2T}\right) .
\label{kohn}
\end{equation}
Therefore, the total non-analytic contribution coming from both the Landau damping and the Kohn anomalies of the particle-hole bubbles can be combined into the following expression:
\begin{eqnarray}
&& \sum_Q \Pi^{\lambda_1 \lambda_2}_{\tau_1 \tau_2}(Q) \Pi^{\lambda_3 \lambda_4}_{\tau_3 \tau_4}(Q)  = \frac{\nu T^3}{4 E_F}  \nonumber \\
&& \times \left[ \mathcal{F}\left(\frac{\Delta^{\lambda_1 \lambda_2}_{\tau_1 \tau_2} + \Delta^{\lambda_3 \lambda_4}_{\tau_3 \tau_4}}{2T}\right) + \mathcal{F}\left(\frac{\Delta^{\lambda_1 \lambda_3}_{\tau_1 \tau_3} + \Delta^{\lambda_2 \lambda_4}_{\tau_2 \tau_4}}{2T}\right)\right] . \label{nonan}
\end{eqnarray}

Using Eq.~(\ref{nonan}) and the expansion (\ref{expansion}) we get the non-analytic parts of the diagrams in Fig.~\ref{fig:diag} at zero temperature $T = 0$:
\begin{eqnarray}
&& \delta \Omega_a = -\frac{\nu v^2}{48 E_F} \sum\limits_{\tau_1, \tau_2} \sum\limits_{\lambda_1, \lambda_2} |\Delta^{\lambda_1 \lambda_2}_{\tau_1, \tau_2}|^3 , \label{oma}\\
&& \delta \Omega_b = 0, \label{omb}\\
&& \delta \Omega_c =  - \frac{\nu u^2}{48 E_F} \sum_{\tau, \lambda_1, \lambda_2}  \label{omc}\\
&&\times  \left[ \frac{b^4}{8 \beta^4}  \left(|\Delta^{\lambda_1 \lambda_1}_{\tau - \tau} + \Delta^{\lambda_2 \lambda_2}_{\tau - \tau}|^3 + |\Delta^{\lambda_1 \lambda_2}_{\tau \tau} + \Delta^{\lambda_1 \lambda_2}_{-\tau - \tau}|^3\right)  \right. \nonumber \\
&& + \left. \frac{\alpha^4}{8 \beta^4}  \left(|\Delta^{\lambda_1 -\lambda_1}_{\tau - \tau} + \Delta^{\lambda_2 -\lambda_2}_{\tau - \tau}|^3 + |\Delta^{\lambda_1 \lambda_2}_{\tau \tau} + \Delta^{-\lambda_1 -\lambda_2}_{-\tau - \tau}|^3\right)  \right., \nonumber\\
&& + \left. \frac{\alpha^2 b^2}{4 \beta^4}  \left(|\Delta^{\lambda_1 \lambda_1}_{\tau - \tau} + \Delta^{\lambda_2 -\lambda_2}_{\tau - \tau}|^3 + |\Delta^{\lambda_1 \lambda_2}_{\tau \tau} + \Delta^{\lambda_1 -\lambda_2}_{-\tau - \tau}|^3\right)  \right], \nonumber\\
&& \delta \Omega_d = \frac{\nu u v}{24 E_F} \sum_{\tau, \lambda} \left[\frac{b^2}{\beta^2} |\Delta^{\lambda\lambda}_{\tau -\tau}|^3 + \frac{\alpha^2}{\beta^2} |\Delta^{\lambda -\lambda}_{\tau - \tau}|^3\right].
\label{omd}
\end{eqnarray}

\begin{figure}[t]
	\includegraphics[width=0.99\columnwidth]{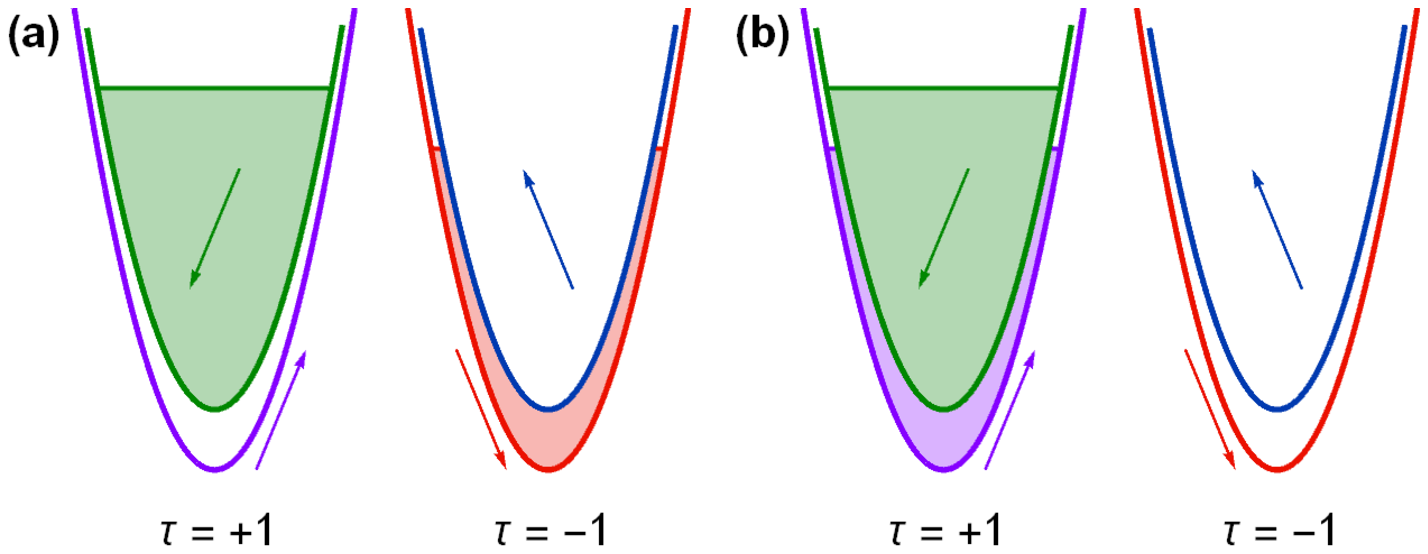}
	\caption{Illustration of the band filling that 
		corresponds to 
		(a) $M_{\beta v} \ne 0$, $M_{\beta} = M_v = 0$;
		(b) $M_v \ne 0$, $M_{\beta} = M_{\beta v} = 0$.
		Example of the band filling corresponding to $M_\beta \ne 0$,
		$M_{\beta v} = M_v = 0$ is shown in Fig.~1 in the main text.}
	\label{fig:ord}
\end{figure}

\section{Legendre transform and free energy}

Here we calculate the free energy per unit area using the Legendre transform:
\begin{eqnarray}
&& F  = \Omega - \sum\limits_{\lambda, \tau} \mu^\lambda_\tau \frac{\partial \Omega}{\partial \mu^\lambda_\tau} = \nonumber \\
&& \Omega + \sum \limits_{\lambda, \tau} \mu^\lambda_\tau n^\lambda_\tau = \Omega + \sum \limits_{\lambda, \tau} \tilde \mu^\lambda_\tau n^\lambda_\tau - \beta \sum\limits_{\lambda, \tau} \lambda n^\lambda_\tau ,
\label{legendre}
\end{eqnarray}
where $n^\lambda_\tau$ is the density of electrons 
corresponding to the band with indices $\lambda$ and $\tau$.
The total density is controlled by external gates and, thus, is constant.
In the main text we define the following order parameters:
\begin{eqnarray}
M_\beta = \sum\limits_{\lambda, \tau} \frac{\lambda n^\lambda_\tau}{4 \nu},\ M_v = \sum\limits_{\lambda, \tau} \frac{\tau n^\lambda_\tau}{4 \nu},\ M_{\beta v} = \sum\limits_{\lambda, \tau} \frac{\lambda \tau n^\lambda_\tau}{4 \nu} .
\label{mag}
\end{eqnarray}
Examples of the filling 
corresponding to non-zero values of one of these order parameters 
are shown in Fig.~1 in the main text 
and in Fig.~\ref{fig:ord}.
The electron densities $n^\lambda_\tau$ can be expressed
in terms of these order parameters
and the total density $n$:
\begin{eqnarray}
&& n^+_+ = \nu \left(M_\beta + M_v + M_{\beta v}\right) + \frac{n}{4} , \\
&& n^+_- = \nu \left(M_\beta - M_v - M_{\beta v}\right) + \frac{n}{4} , \\
&& n^-_+ = \nu \left(-M_\beta + M_v - M_{\beta v}\right) + \frac{n}{4} , \\
&& n^-_- = \nu \left(-M_\beta - M_v + M_{\beta v}\right) + \frac{n}{4} .
\end{eqnarray}
Since the order parameters are assumed to be much smaller than $E_F$, the densities $n^\lambda_\tau$ are all well-defined and non-negative.
We use these relations to calculate the following sums:
\begin{eqnarray}
&& \sum\limits_{\lambda, \tau} \left(n^\lambda_\tau \right)^2 = \frac{n^2}{4} + 4 \nu^2 \left(M_\beta^2 + M_v^2 + M_{\beta v}^2\right) , \label{nn} \\
&& \sum\limits_{\lambda, \tau} n^\lambda_\tau n^{\lambda}_{-\tau} = \frac{n^2}{4} + 4 \nu^2 \left(M_\beta^2 - M_v^2 - M_{\beta v}^2\right) , \label{nnb}\\
&& \sum\limits_{\lambda, \tau} n^\lambda_\tau n^{-\lambda}_{\tau} = \frac{n^2}{4} + 4 \nu^2 \left(-M_\beta^2 + M_v^2 - M_{\beta v}^2\right) , \label{nnv} \\
&& \sum\limits_{\lambda, \tau} n^\lambda_\tau n^{-\lambda}_{-\tau} = \frac{n^2}{4} + 4 \nu^2 \left(-M_\beta^2 - M_v^2 + M_{\beta v}^2\right) , \label{nnbv}\\
&& \sum\limits_{\lambda, \tau} \left|n^\lambda_\tau - n^{-\lambda}_\tau \right|^3 = 32 \nu^3 f(M_\beta, M_{\beta v}) , \label{s1}\\
&& \sum\limits_{\lambda, \tau} \left|n^\lambda_\tau - n^{\lambda}_{-\tau} \right|^3 = 32 \nu^3 f(M_v, M_{\beta v}) , \label{s2}\\
&& \sum\limits_{\lambda, \tau} \left|n^\lambda_\tau - n^{-\lambda}_{-\tau} \right|^3 = 32 \nu^3 f(M_\beta, M_{v}) , \label{s3}
\end{eqnarray}
where we introduced the following symmetric function:
\begin{eqnarray}
&& f(x, y) = \frac{1}{2} \left(|x + y|^3 + |x - y|^3\right) .
\label{f}
\end{eqnarray}

We calculated the grand canonical potential
within the SCB approximation, see Eq.~(\ref{omscb}),
and also included small cubic corrections 
coming from the dynamical screening of the Coulomb interaction and
the interaction vertex correction,
see Eqs.~(\ref{oma})--(\ref{omd}):
\begin{eqnarray}
&& \Omega = \Omega^{SCB} + \delta \Omega, \label{omtot}\\
&& \delta\Omega = \delta \Omega_a + \delta \Omega_b + \delta \Omega_c + \delta \Omega_d , \label{delom}
\end{eqnarray}
where $\delta \Omega$ is the total cubic correction.

Let us first set $\delta \Omega$ to zero and calculate 
the SCB contribution to the free energy.
The electron densities are given by 
derivatives of $\Omega^{SCB}$, see Eq.~(\ref{omscb}),
with respect to the chemical potentials:
\begin{eqnarray}
&& n^\lambda_\tau  = - \frac{\partial \Omega^{SCB}}{\partial \mu^\lambda_\tau} 
= \frac{\nu}{(1 - v)^2 - u^2} \left[(1 - v) \tilde \mu^\lambda_\tau  \right. \nonumber \\
&& \left. +  u \left(\frac{b^2}{\beta^2} \tilde \mu^\lambda_{-\tau} + \frac{\alpha^2}{\beta^2} \tilde \mu^{-\lambda}_{-\tau}\right)\right] . \label{denmu}
\end{eqnarray}
This is the linear relation which can be inverted exactly:
\begin{widetext}
	\begin{eqnarray}
	&& \nu \tilde \mu^\lambda_\tau = \frac{1 - v}{(1 - v)^2 - \zeta^2 u^2} \left\{\left[(1 -v)^2 - \frac{u^2}{2} (1 +\zeta^2)\right] n^\lambda_\tau + \frac{u^2}{2} (1 - \zeta^2) n^{-\lambda}_\tau \right\} \nonumber \\
	&& - \frac{u}{(1 - v)^2 - \zeta^2 u^2} \left\{ \frac{1 + \zeta }{2} \left[(1 - v)^2 - \zeta u^2\right] n^\lambda_{-\tau} + \frac{1 - \zeta }{2} \left[(1 - v)^2 + \zeta u^2\right] n^{-\lambda}_{-\tau} \right\} ,
	\label{bigmu}
	\end{eqnarray}
\end{widetext}
where 
\begin{eqnarray}
&& \zeta = \frac{b^2 - \alpha^2}{\beta^2} = \frac{b^2 - \alpha^2}{b^2 + \alpha^2} .
\end{eqnarray}
Here we assume $u \ll v$ and expand 
Eq.~(\ref{bigmu}) with respect to $u$
leaving only the first order terms in $u$,
while $v$ is treated non-perturbatively:
\begin{align}
\nu \tilde \mu^\lambda_\tau \approx (1 - v) n^\lambda_\tau - \frac{u}{2} \left[(1 + \zeta) n^\lambda_{-\tau} + (1 - \zeta) n^{-\lambda}_{-\tau}\right] . \label{mun}
\end{align}
The SCB contribution to the free energy then
reads:
\begin{eqnarray}
&& F^{SCB} = \Omega^{SCB} + \sum\limits_{\lambda, \tau} \tilde \mu^\lambda_\tau n^\lambda_\tau - 4 \nu \beta M_\beta = \nonumber \\
&& \frac{1}{2} \sum\limits_{\lambda, \tau} \tilde \mu^\lambda_\tau n^\lambda_\tau - 4 \nu \beta M_\beta .
\label{FSCB}
\end{eqnarray}
Here we used the following identity:
\begin{eqnarray}
&& \sum\limits_{\lambda, \tau} \tilde \mu^\lambda_\tau n^\lambda_\tau = - 2 \Omega^{SCB}, \label{simp}
\end{eqnarray}
where the factor $2$
comes from the quadratic scaling of $\Omega^{SCB}$ 
with the chemical potentials.
Substituting the chemical potentials
given by Eq.~(\ref{mun})
into Eq.~(\ref{FSCB}),
we find the SCB contribution to the free energy:
\begin{eqnarray}
&& F^{SCB} = \frac{1 - v}{2 \nu} \sum\limits_{\lambda, \tau} \left(n^\lambda_\tau\right)^2 - 4 \nu \beta M_\beta \nonumber \\
&& - \frac{u}{4 \nu} \sum\limits_{\lambda, \tau} \left[(1 + \zeta) n^\lambda_\tau n^\lambda_{-\tau} + (1 - \zeta) n^\lambda_\tau n^{-\lambda}_{-\tau}\right] .
\end{eqnarray}
Finally, we use the sums given in Eqs.~(\ref{nn})--(\ref{nnbv})
in order to represent the free energy
in terms of the order parameters
given by Eq.~(\ref{mag}).
This gives Eq.~(8) in the main text.

Next, we calculate the cubic correction to the free energy.
From Eq.~(\ref{legendre}), we find the correction  to the free energy:
\begin{eqnarray}
&& \delta F = \delta \Omega + \delta \Omega^{SCB} +  \sum\limits_{\lambda, \tau} \delta\tilde \mu^\lambda_\tau n^\lambda_\tau ,
\label{df}
\end{eqnarray}
where $\delta \Omega$ is the cubic correction, see Eq.~(\ref{delom}),
$\delta \Omega^{SCB}$ comes from the change in the chemical 
potentials:
\begin{eqnarray}
&& \delta \Omega^{SCB} = \sum\limits_{\lambda, \tau} \frac{\partial \Omega^{SCB}}{\partial \mu^\lambda_\tau} \delta \tilde \mu^\lambda_\tau = \nonumber \\
&& -\sum\limits_{\lambda, \tau} \left(n^\lambda_\tau - \delta n^\lambda_\tau\right) \delta \tilde \mu^\lambda_\tau \approx
- \sum\limits_{\lambda, \tau} n^\lambda_\tau \delta \tilde \mu^\lambda_\tau , \label{domscb}
\end{eqnarray}
where we used that $\Omega^{SCB} = \Omega - \delta \Omega$,
see Eq.~(\ref{omtot}),
$\delta n^\lambda_\tau = - \partial \delta \Omega/\partial \mu^\lambda_\tau$
is the non-analytic contribution to the densities.
Substituting Eq.~(\ref{domscb}) back into Eq.~(\ref{df}),
we get the non-analytic correction to the free energy:
\begin{eqnarray}
&& \delta F = \delta \Omega (\tilde \mu^\lambda_\tau = n^\lambda_\tau / \nu) , \label{cubf}
\end{eqnarray}
where we used the relation $\tilde \mu^\lambda_\tau = n^\lambda_\tau / \nu$ 
that is true in the non-interacting 2DEG
because the cubic correction $\delta \Omega$,
see Eq.~(\ref{delom}),
was calculated within second order perturbation theory.
Including the interaction corrections in $\tilde \mu^\lambda_\tau$
is beyond the applicability of the approximation
that we used.
As we neglected the $u^2$ terms in the SCB approximation,
see Eq.~(\ref{mun}),
we have to do the same with the cubic correction,
i.e. we neglect $\delta \Omega_c$, see Eq.~(\ref{omc}).
Thus, the cubic correction is given
by Eqs.~(\ref{oma}), (\ref{omd})
and can be simplified using $\tilde \mu^\lambda_\tau = n^\lambda_\tau / \nu$ 
and the sums given by Eqs.~(\ref{s1})--(\ref{s3}).
This results in Eq.~(16) in the main text.


\begin{thebibliography}{10}
	\bibitem{dress} M. Dresselhaus, G. Dresselhaus, S. B. Cronin, A. G. S. Filho, Solid State Properties: From Bulk to
	Nano (Springer, Berlin, Heidelberg, ed. 1, 2018).
	
	\bibitem{awschalom} D. D. Awschalom, D. Loss, and N. Samarth (ed.),
	Semiconductor Spintronics And Quantum Computation (Springer Science and Business Media, 2013).
	
	\bibitem{ando}  T. Ando, A. B. Fowler, F. Stern, Rev. Mod. Phys. \textbf{54}, 437 (1982).
	
	\bibitem{miserdot} D. Miserev, and O. P. Sushkov, Phys. Rev. B \textbf{100}, 205129 (2019).
	
	\bibitem{ashcroft} N. W. Ashcroft and N. D. Mermin, Solid State Physics (Saunders, New York, 1974).
	
	\bibitem{wang} Q. H. Wang, K. Kalantar-Zadeh, A. Kis, J. N. Coleman, and M. S. Strano, Nat. Nanotechnol. \textbf{7}, 699--712 (2012).
	
	\bibitem{xu} X. Xu, W. Yao, D. Xiao, and T. F. Heinz,
	Nat. Phys. \textbf{10}, 343--350 (2014).
	
	
	
	
	\bibitem{mukh} D. K. Mukherjee, A. Kundu, and H. A. Fertig,
	Phys. Rev. B \textbf{98}, 184413 (2018).
	
	\bibitem{braz} J. E. H. Braz, B. Amorim, and E. V. Castro,
	Phys. Rev. B \textbf{98}, 161406(R) (2018).
	
	\bibitem{donck} M. Van der Donck and F. M. Peeters, 
	Phys. Rev. B \textbf{98}, 115432 (2018).
	
	\bibitem{miserfer} D. Miserev, J. Klinovaja, and D. Loss, Phys. Rev. B \textbf{100}, 014428 (2019).
	
	
	
	
	\bibitem{rama} A. Ramasubramaniam, Phys. Rev. B \textbf{86}, 115409 (2012).
	
	\bibitem{chei} T. Cheiwchanchamnangij and W. R. L. Lambrecht,
	Phys. Rev. B \textbf{85}, 205302 (2012).
	
	\bibitem{zhao} W. Zhao, Z. Ghorannevis, L. Chu, M. Toh, C. Kloc, P.-H. Tan, and G. Eda, 
	ACS Nano \textbf{7}, 791--797 (2012).
	
	\bibitem{ross} J. S. Ross, S. Wu, H. Yu, N. J. Ghimire, 
	A. M. Jones, G. Aivazian, J. Yan, D. G. Mandrus, D. Xiao, 
	W. Yao, and X. Xu, 
	Nat. Commun. \textbf{4}, 1474 (2013).
	
	
	\bibitem{kadantsev} E. S. Kadantsev and P. Hawrylak,
	Solid State Commun. \textbf{152}, 909–913 (2012).
	
	\bibitem{xiao} D. Xiao, G.-B. Liu, W. Feng, X. Xu, and W. Yao, Phys. Rev. Lett. \textbf{108}, 196802 (2012).
	
	\bibitem{kosmider} K. Ko\'{s}mider, J. W. Gonz\'{a}lez, and J. Fern\'{a}ndez-Rossier,
	Phys. Rev. B \textbf{88}, 245436 (2013).
	
	\bibitem{liu} G.-B. Liu, W.-Y. Shan, Y. Yao, W. Yao, and D. Xiao,
	Phys. Rev. B \textbf{88}, 085433 (2013).
	
	\bibitem{kli}  J. Klinovaja and D. Loss, Phys. Rev. B {\bf 88}, 075404 (2013).
	
	\bibitem{korma} A. Korm\'{a}nyos, V. Z\'{o}lyomi, N. D. Drummond, and G. Burkard, Phys. Rev. X \textbf{4}, 011034 (2014).
	
	\bibitem{burkard} A. Korm\'{a}nyos, G. Burkard, M. Gmitra, J. Fabian, V. Z\'{o}lyomi, N. D. Drummond, and V. Fal'ko,
	2D Mater. \textbf{2}, 022001 (2015).
		
	
	
	
	\bibitem{bkv} D. Belitz, T. R. Kirkpatrick, and T. Vojta, Rev. Mod. Phys. \textbf{77}, 579 (2005).
	
	
	\bibitem{roch} J. G. Roch, G. Froehlicher, N. Leisgang, P. Makk, K. Watanabe, T. Taniguchi, and R. J. Warburton, Nat. Nanotechnol. \textbf{14}, 432--436 (2019).
	
	\bibitem{rochfirst} J. G. Roch, D. Miserev, G. Froehlicher, N. Leisgang, L. Sponfeldner,
	K. Watanabe, T. Taniguchi, J. Klinovaja, D. Loss, and R. J. Warburton, Phys. Rev. Lett. \textbf{124}, 187602 (2020).
	
	\bibitem{pisoni} R. Pisoni, A. Korm\'{a}nyos, M. Brooks, 
	Z. Lei, P. Back, M. Eich, H. Overweg, 
	Y. Lee, P. Rickhaus, K. Watanabe, T. Taniguchi, A. Imamoglu,
	G. Burkard, T. Ihn, and K. Ensslin,
	Phys. Rev. Lett. \textbf{121}, 247701 (2018).
	
	
	
	
	\bibitem{belitz} D. Belitz and T. R. Kirkpatrick, Phys. Rev. Lett. \textbf{89}, 247202 (2002).
	
	\bibitem{maslov} D. L. Maslov and A. V. Chubukov, Phys. Rev. B \textbf{79}, 075112 (2009).

	\bibitem{kirk} T. R. Kirkpatrick and D. Belitz, Phys. Rev. B
	\textbf{67}, 024419 (2003).
	
	\bibitem{zak1} R. A. Zak, D. L. Maslov, and D. Loss, Phys. Rev. B \textbf{82}, 115415 (2010).
	
	\bibitem{zak2} R. A. Zak, D. L. Maslov, and D. Loss, Phys. Rev. B \textbf{85}, 115424 (2012).
	
	\bibitem{kb} T. R. Kirkpatrick and D. Belitz, Phys. Rev. Lett. \textbf{124}, 147201 (2020).
	
	
	\bibitem{stoner} E. C. Stoner, Proc. R. Soc. Lond. A \textbf{165}, 372--414 (1938).
	
	
	\bibitem{guinea} R. Roldan, E. Cappelluti, and F. Guinea, Phys. Rev. B \textbf{88}, 054515 (2013).
	
	
    \bibitem{SM} 
	See Supplemental Material at . . . for the derivation of the free energy 
	including the interaction vertex correction and the dynamical screening of the Coulomb interaction.
	
	
	\bibitem{pt} D. Belitz, T. R. Kirkpatrick, and T. Vojta,
	Phys. Rev. Lett. \textbf{82}, 4707 (1999).
	
	\bibitem{pepin} A. V. Chubukov, C. P\'{e}pin, and J. Rech,
	Phys. Rev. Lett. \textbf{92}, 147003 (2004).
	
	\bibitem{sachdev} S. Sachdev, Quantum Phase Transitions (Cambridge
	University Press, ed. 2, 2011).
	
	\bibitem{mvojta} H. v. L\"{o}hneysen, A. Rosch, M. Vojta, and P. W\"{o}lfle,
	Rev. Mod. Phys. \textbf{79}, 1015 (2007).
	
	\bibitem{abrikosov} A. A. Abrikosov, L. P. Gorkov, and I. E. Dzialoshinskii,
	Quantum Field Theoretical Methods in Statistical Physics
	(Pergamon, 1965).
	
	
\end{thebibliography}
\end{document}